\documentclass[aps,prb,twocolumn,superscriptaddress]{revtex4-1}
\usepackage{amsmath}
\usepackage{amsfonts}
\usepackage{amssymb}
\usepackage{graphicx}
\usepackage{color}
\usepackage{xcolor}
\usepackage{ulem}  
\usepackage{mathtools}

\begin{document}

\title{Topological theory of inversion-breaking charge-density-wave monolayer 1\textit{T}-TiSe$_2$}

\author{Ming-Chien Hsu}
\affiliation{Department of Physics, National Sun Yat-sen University, Kaohsiung 80424, Taiwan}

\author{Bahadur Singh}
%\email{bahadursingh24@gmail.com}
%\affiliation{SZU-NUS Collaborative Center and International Collaborative Laboratory of 2D Materials for Optoelectronic Science $\&$ Technology, Engineering Technology Research Center for 2D Materials Information Functional Devices and Systems of Guangdong Province, College of Optoelectronic Engineering, Shenzhen University, ShenZhen 518060, China}
%\affiliation{Department of Physics, Northeastern University, Boston, Massachusetts 02115, USA}
\affiliation{Department of Condensed Matter Physics and Materials Science, Tata Institute of Fundamental Research, Colaba, Mumbai 400005, India}

\author{Chuang-Han Hsu}
%\affiliation {Centre for Advanced 2D Materials and Graphene Research Centre, National University of Singapore, 6 Science Drive 2, Singapore 117546}
%\affiliation {Department of Physics, National University of Singapore, 2 Science Drive 3, Singapore 117542}
\affiliation{Department of Electrical and Computing Engineering, National University of Singapore, Singapore 117583, Singapore}

\author{Su-Yang~Xu}
%\affiliation{Department of Physics, Massachusetts Institute of Technology, Cambridge, Massachusetts 02139, USA}
\affiliation{Department of Chemistry and Chemical Biology, Harvard University, Cambridge, Massachusetts 02138, USA}

\author{Hsin~Lin}
\affiliation {Institute of Physics, Academia Sinica, Nankang Taipei 11529, Taiwan}
\date{\today }

\author{Shin-Ming Huang}
\email{shinming@mail.nsysu.edu.tw}
\affiliation{Department of Physics, National Sun Yat-sen University, Kaohsiung 80424,
Taiwan}

\begin{abstract}
A charge density wave (CDW) of a nonzero ordering vector $\mathbf{q}$ couple electronic states at $\mathbf{k}$ and $\mathbf{k}+\mathbf{q}$ statically, giving rise to a reduced Brillouin zone (RBZ) due to the band folding effect. Its structure, referred to an irreducible representation of the little group of $\mathbf{q}$, would change the symmetry of the system and electronic structure accompanying possible change of band inversion,
offering a chance of the topological phase transition. Monolayer 1\textit{T}-TiSe$_2$ is investigated for it shows an unconventional CDW phase having a triple-$q$ $M_1^-$ structure.
Moreover, the coupling between the triple-$q$ component of the $M_1^-$ CDW will inevitably produces a small $M_1^+$ CDW.  
The CDW yields a band inversion in 1\textit{T}-TiSe$_2$ but different types of CDW can affect the electronic structure and system topology differently.
The impact of CDW of different types was studied by utilizing a symmetrization-antisymmetrization technique to extract the $M_1^-$ and $M_1^+$ CDW contributions in the DFT-based tight-binding model and study their effects.
The results are consistent with the parity consideration, improving understanding of topology for a CDW system with and without parity.
\end{abstract}

\maketitle
\section{Introduction}
Since the discovery of topological insulators (TI) \cite{TI_Kane_Moore, Hasan_Kane10, Ando13}, much attention has been paid to figure out the topological nature, characterized by the $\mathbb{Z}_2$ invariant, in time reversal symmetric systems.
The topological phase transition due to change of $\mathbb{Z}_2$ can be achieved by several methods like tuning some parameters in the system \cite{Murakami06_TopPT}, for example applying the stress \cite{strain_Young11, honeycomb_Nie15, MX2_Lin17}.
It was recently discussed that the topological state can also be changed by the charge density waves (CDW), through huge modification of the electronic structure, resulting in dissolution of corresponding band inversion \cite{CDW20_VTe2}.

In the CDW system, the translational symmetry is broken due to charge modulated structure.
For commensurate CDW, the symmetry can be restored with multiple times of the direct basis translation of the normal (undistorted) phase, leading to an enlarged unit cell, and hence described by a reduced Brillouin zone (RBZ) in the reciprocal space.
The RBZ results in band folding, which basically identifies two different $\mathbf{k}$'s points in the primitive BZ. 
When the indirect gap (or overlap) with momentum difference can be coupled to make the folded conduction and valence bands overlap, it results in band inversion between distinct $\mathbf{k}$ points, different to the common band inversion at a particular $\mathbf{k}$ point.
The coupling between different $\mathbf{k}$ points to produce topological transition, if achieved, would be interesting and may provide insights to have more ways to control the topological transitions.

The vector connecting the two identified $\mathbf{k}$'s is the ordering vector $\mathbf{q}$.
The CDW can be described by a single $\mathbf{q}$ or multiple $\mathbf{q}$'s altogether.
The system is then turned by the CDW into a phase with space group $\mathcal{G}$, and the order parameter (OP) $\psi$ of the CDW now has the symmetry described by the little group of $\mathbf{q}$, a subgroup of the primitive space group $\mathcal{G}_0$ whose symmetry elements leave $\mathbf{q}$ invariant.
There can be different representations associated with the little group, resulting in the CDW of distinct types.
For example, starting from the inversion-symmetric primitive phase, a developed OP $\psi$ of the CDW can have a one-dimensional representation such that $I: \psi \mapsto \pm \psi$, where $I$ is the inversion operation.
Interestingly, when the minus sign $-1$ is taken, the CDW system has its inversion symmetry broken.
Different representations of the OP result in the gap function $\Delta$ of distinct types coupling $\mathbf{k}$ and $\mathbf{k}+\mathbf{q}$.
The coupling by the gap function $\Delta$, accompanying band reordering from folding, can modify the electronic structure and potentially change the system topology.
However, the gap function $\Delta$ of different types is seldom discussed in topological effects, to the best of our knowledge. We will address this issue in this article.

%\subsection{Gamma M coupling}
The material 1\textit{T}-TiSe$_2$ has been long known to have a low-temperature CDW phase.
The origin of the CDW is still controversial but may be caused by the Coulomb interaction induced excitonic condensate \cite{CDW15_MBE, CDW16_ARPES, Exciton_CDW14, Singh_CDW17, Singh_CDW18}
 or electron-phonon coupling \cite{Exciton_phonon11, Exciton_phonon18}.
Interestingly, the signature of inversion symmetry breaking in (3D) bulk 1\textit{T}-TiSe$_2$ with CDW was observed \cite{Su_TiSe2}.
Our previous work has gone through symmetry as well as topology thoroughly in bulk 1\textit{T}-TiSe$_2$ \cite{huang2018}. 
Since the combined operation of inversion with translation along the vertical $z$ axis remains a generalized inversion, which also corresponds to an inversion with respect to a shifted center at $(0, 0, \frac{1}{2})$, the technique of $\mathbb{Z}_2$ calculation using products of parity eigenvalues at time-reversal invariant momentum (TRIM) points applies \cite{Fu07_Inv}.

\begin{figure}
\begin{center}
\includegraphics[width=0.5\textwidth]{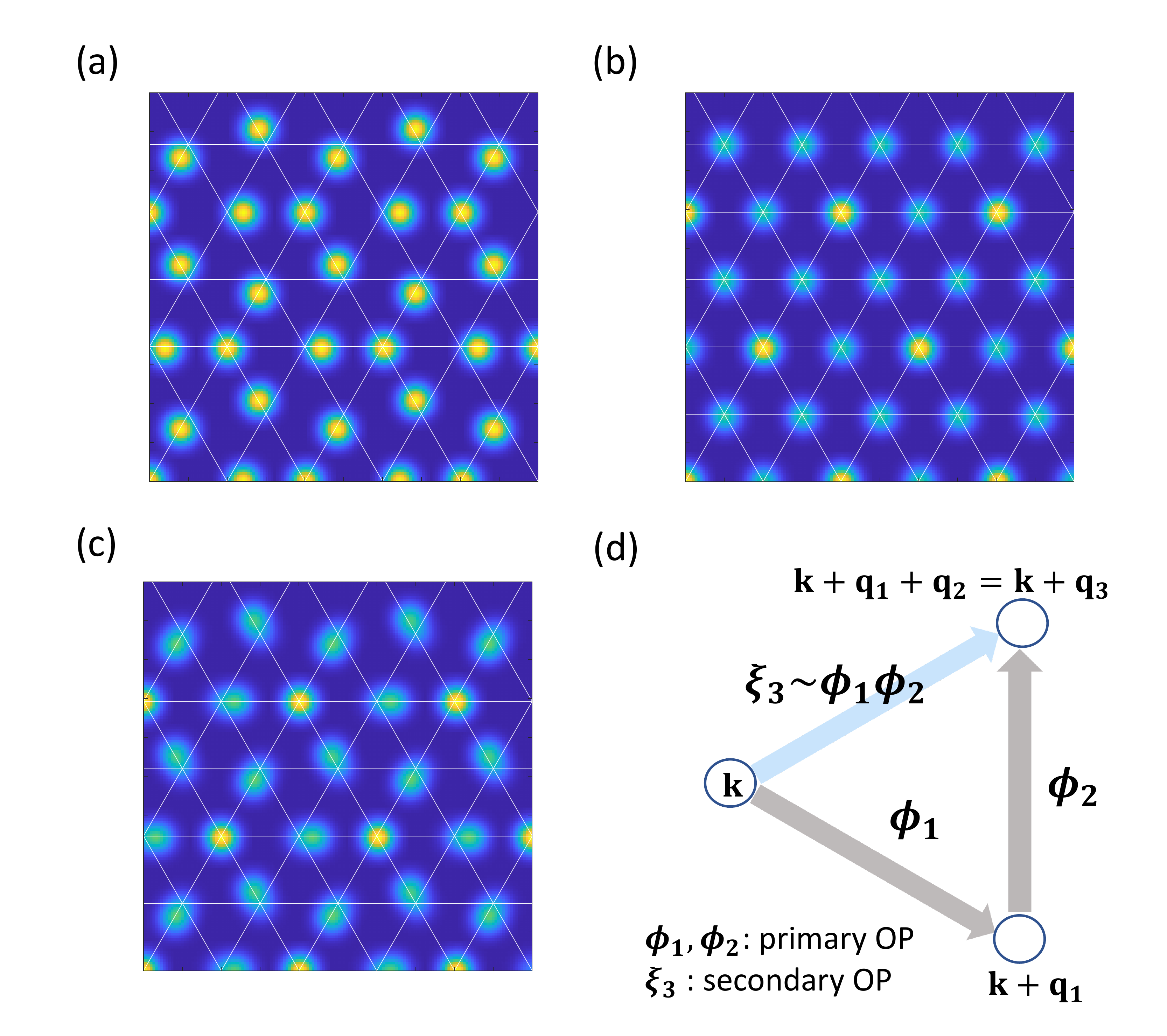}
\caption{
(a)-(c) illustrate triple-$q$ CDWs in a triangular lattice. They are $M_1^-$ (a), $M_1^+$ (b), and the combination of $M_1^-$ and $M_1^+$ CDWs (c), respectively. In the illustration, no sublattice is considered and one can regard the profile as the charge from Ti atoms. 
(d) Successive couplings to two primary $M_1^-$ OPs ($\phi_1,~\phi_2$) get the same effect as a coupling to a secondary $M_1^+$ OP ($\xi_3$). 
\label{fig:1}}
\end{center}
\end{figure}

Recently the monolayer system has been also found to have a robust CDW phase with higher transition temperature \cite{CDW15_MBE, CDW16_ARPES, CDW12_Raman, CDW15_MLrobust, Singh_CDW17}. 
Different to the bulk case, the inversion symmetry cannot be restored via translation along the $z$ axis. Here 
the ordering vector $\mathbf{q}$ connecting from $\Gamma$ to $M$ has the little cogroup $C_{2h}$ and it restricts the OP to be one of the four possible irreducible representations (IRs) $\{M_1^+, M_1^-, M_2^+, M_2^-\}$ (Table\ \ref{tab:C2h}), among which the energy favourable triple-$q$ $M_1^-$ $2\times 2$ CDW breaking inversion symmetry 
 is the case of interest \cite{Exciton_phonon18}. 
 The triple-$q$ indicates that three ordering vectors, differing by $\pm120^{\circ}$, coexist to maintain the threefold rotation symmetry. 
The superscript $+(-)$ refers to even (odd) parity of the OP, that is, the eigenvalue of inversion operation (Table\ \ref{tab:C2h}).  
Figure~\ref{fig:1} demonstrates a triple-$q$ $M_1^-$ CDW (a), a triple-$q$ $M_1^+$ CDW (b), and their mixture (c). Some charge centers in $M_1^+$ deviate from the lattice points, reflecting the oddness property. 
Because of symmetry, the combined operation of any two components of $M_1^-$ OP will inevitably result in an $M_1^+$ OP effectively, Fig. \ref{fig:1}(d). As the result, the system has coexistence $M_1^-$ and $M_1^+$ CDWs.
Since the $M_1^+$ and $M_1^-$ OPs have opposite parities, they are expected to give rise to different topologies in the electronic system.
This mixture in monolayer 1\textit{T}-TiSe$_2$ does not happen in the bulk~\cite{huang2018}, motivating us to utilize a symmetrization approach to disclose the symmetry and topology in this work.

The organization of the article is as follows. 
The structures and {\it ab initio} calculations for the primitive (normal) and CDW 1\textit{T}-TiSe$_2$  are given in Sec.\ \ref{Sec:DFTband}. The coexistence of the $M_1^-$ and $M_1^+$ OP representations from the Ginzburg-Landau (GL) energy consideration is shown in Sec.\ \ref{Sec:GLeng}. 
The CDW coupling perturbation is symmetrized and anti-symmetrized by the inversion to study the corresponding topological effects in Sec.\ \ref{Sec:Mixing_TB}. 
A topological phase transition can happen with varying mixed strengths of inversion (parity-) even and odd perturbations in the band-folded system, which is discussed in Sec.\ \ref{Sec:discussion}.

\section{Crystal and electronic structures \label{Sec:DFTband}}
\begin{figure}
\begin{center}
\includegraphics[width=0.5\textwidth]{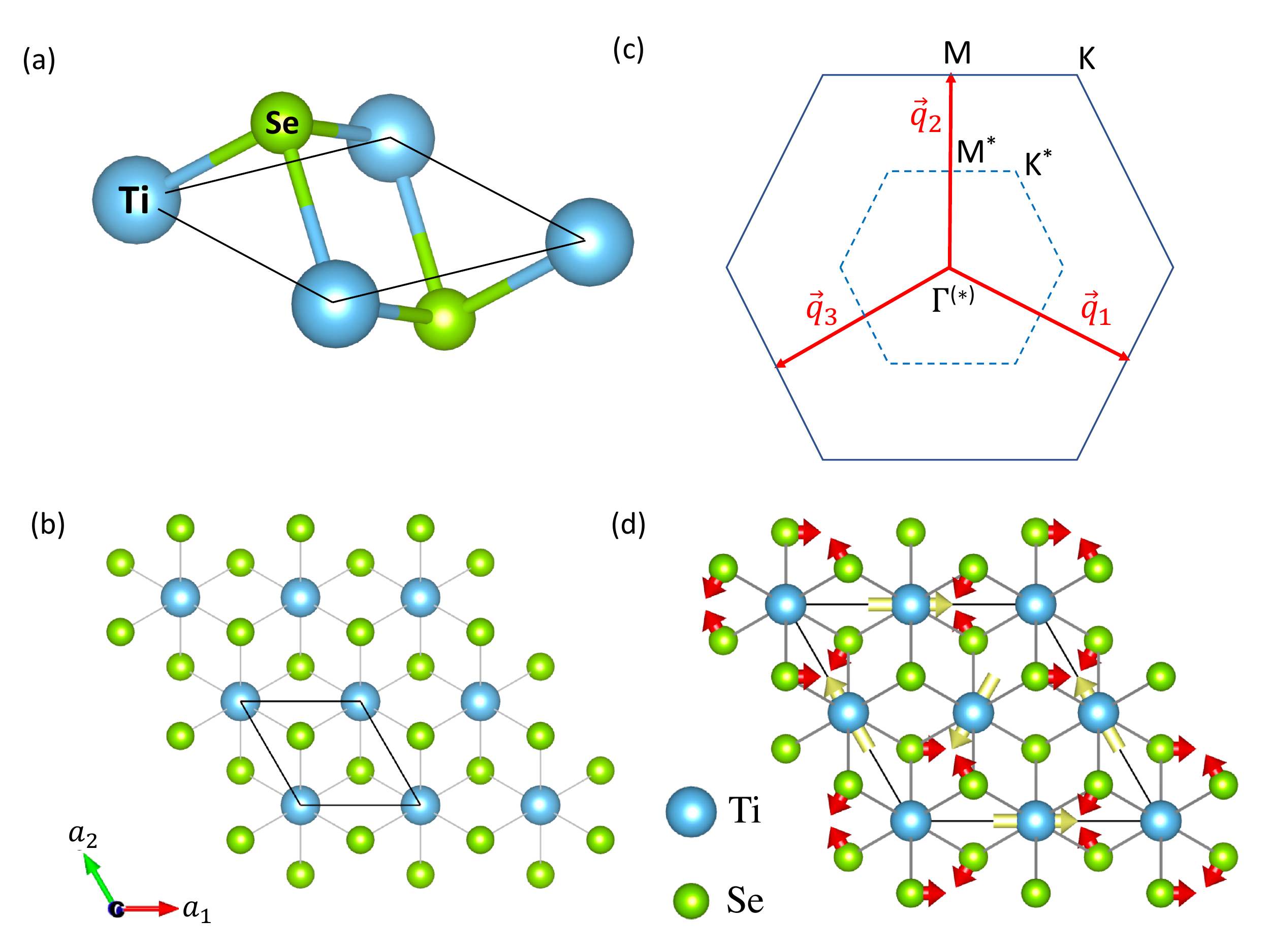}
\caption{
(a) The unit cell of the primitive monolayer TiSe$_2$ is shown, and its top view is shown in (b).
(c) The large hexagon with solid line represents the Brillouin zone (BZ) of the primitive cell.
The red lines $\mathbf{q}_1$, $\mathbf{q}_2$, and $\mathbf{q}_3$ connecting $\Gamma$ and each $M$ are ordering vectors.
The small hexagon with dashed line is the reduced BZ (RBZ) for the $2\times 2$ CDW supercell. The labels with asterisk are for symmetry points in a CDW phase, for example, $\Gamma^{*}$ and $M^{*}$. 
(d) The top view of the atomic structure for the TiSe$_2$ in $M_1^-$ CDW phase is shown. The arrow represents the direction of the ionic displacement.
\label{fig:2}}
\end{center}
\end{figure}

Pristine monolayer 1$T$-TiSe$_{2}$ has a triangular lattice with the Ti layer sandwiched by two Se layers. A Ti atom is surrounded by six Se atoms, three on the top and three at the bottom [Fig.\ \ref{fig:2} (a) and (b)]. 
The in-plane lattice constant is 3.54 \AA~for nearest neighbor Ti atoms. Se atoms are located above and below the plane with height of 1.54 \AA. The corresponding hexagonal BZ and relevant symmetry points are shown in solid lines of Fig.\ \ref{fig:2} (c).
The space group is $P\bar{3}m1$ (no. 164), whose point group contain a threefold rotation axis ($3_{\left[001\right]}$), three conjugate twofold axes ($2_{\left[100 \right]}$, $2_{\left[010 \right]}$, $2_{\left[110 \right]}$), three conjugate mirror planes ($m_{\left[100 \right]}$, $m_{\left[010 \right]}$, $m_{\left[110 \right]}$), and an inversion center ($i$; there are multiple choices of the inversion center, either at a Ti site or at the middle point of two nearest Ti atoms). 

When the system enters the CDW phase, a $2\times 2$ superstructure is present. We illustrate the CDW by the lattice distortion as arrows for atoms' displacements in Fig.\ \ref{fig:2} (d). Referring to Fig.\ \ref{fig:2} (d), the CDW phase preserves the perpendicular threefold rotation axis as well as the in-plain twofold axes, while it breaks mirror planes and the inversion center. Therefore, the space group changes to $P321$ (no. 150). Meanwhile, the fourfold extension of the unit cell indicates a RBZ as shown by dashed lines in Fig.\ \ref{fig:2} (c). In the presence of a CDW with ordering vector $\mathbf{q}$, momenta $\mathbf{k}$ and $\mathbf{k}+\mathbf{q}$ in the normal state become equivalent in the CDW state. Therefore, $\Gamma$ and $M$s will be identified as $\Gamma^*$ in the RBZ. (To make distinction, symmetry-point labels with subscript $*$ are for the CDW state.)

We performed the band structure calculations within the density functional theory (DFT) using the Vienna Ab initio Simulation Package (VASP) \cite{VASP93, VASP_JCMS96, VASP_PRB96} on both the normal and CDW phases. The spin-orbit coupling (SOC) is considered.
Lattice relaxation till the residual forces on each ion less than $10^{-3}$ eV/\AA~is adopted to optimize the lattice parameters and atomic positions. 
We executed both GGA (generalized gradient approximation) \cite{PBE96} and HSE (Heyd-Scuseria-Ernzerhof) \cite{HSE_JCP03} approximations. We observed that the GGA result showed too much energy overlap among the valence bands and conduction band~\cite{Singh_CDW17}.
To incorporate more correlation effects to expel energy overlap, we adopted the HSE approximation for the exchange-correlation functional.
After the simulations, we used the wannierization method to obtain the hopping integrals for tight-binding model (TB) for both the normal state ($H_0$) and the CDW state ($H$). With the hopping integrals, we will tune the TB to study the effects of the CDW on band structures and topology.

\begin{figure}
\begin{center}
\includegraphics[width=0.5\textwidth]{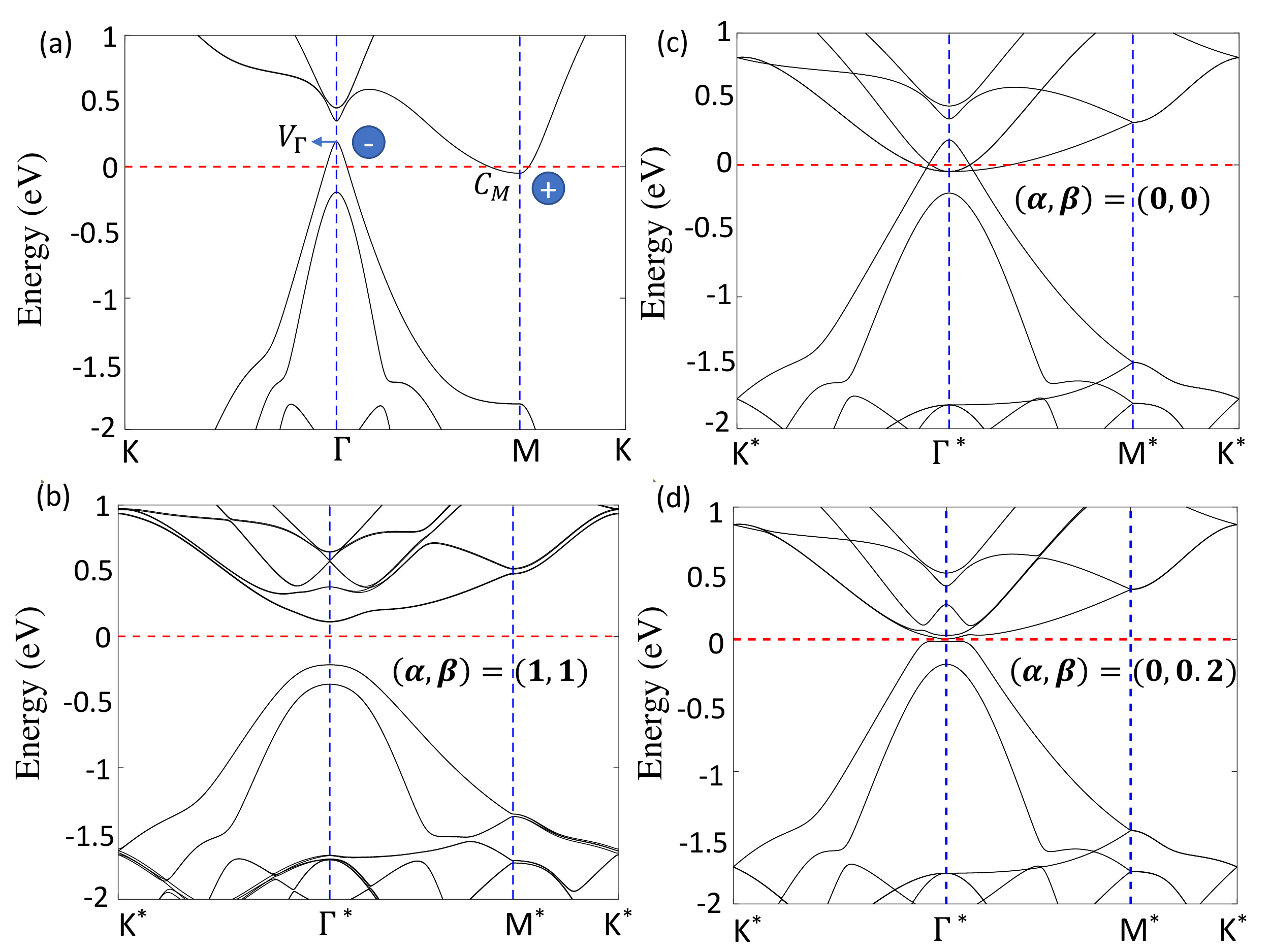}
\caption{All the band structures shown are results with spin-orbit coupling (SOC) calculated under the HSE approximation.
The band structure for (a) the normal-state phase, (b) the CDW phase with $(\alpha, \beta) = (1, 1)$,  (c) the primitive phase in the $2\times 2$ supercell, $(\alpha, \beta) = (0, 0)$, and
(d) the system with a purely parity-odd perturbation, $(\alpha, \beta) = (0, 0.2)$. In (a), the parity of the top valence band at $\Gamma$ is $-1$, and that of the lowest conduction band at $M$ is $+1$.
\label{fig:3}}
\end{center}
\end{figure}

We show the DFT-HSE band structures of monolayer 1$T$-TiSe$_{2}$ with SOC for the normal state (primitive) in Fig.\ \ref{fig:3} (a) and the CDW phase in Fig.\ \ref{fig:3} (b), respectively. 
In the normal state, Fig.\ \ref{fig:3} (a), one can see a semimetallic band structure that a hole pocket at $\Gamma$ and an electron pocket at $M$ (there are another two at other $M$s) have an appreciable energy overlap. The energy overlap might be relevant because the CDW order will couple these pockets via band folding.
The folding will give band inversion with a CDW gap, resulting in the final electronic structure shown in Fig.\ \ref{fig:3} (b).

% ------------------------- %
%              figure                    %
% ------------------------- %

\begin{table}[tbp]
\begin{center}
\setlength{\tabcolsep}{1.5em}
\begin{tabular}{c|cccc}
\hline
$C_{2h}$ & $E$ & $C_{2}$ & $\sigma _{h}$ & $i$ \\ \hline
$M_{1}^{+}$ ($A_{g}$) & $1$ & $1$ & $1$ & $1$ \\ 
$M_{2}^{+}$ ($B_{g}$) & $1$ & $-1$ & $-1$ & $1$ \\ 
$M_{1}^{-}$ ($A_{u}$) & $1$ & $1$ & $-1$ & $-1$ \\ 
$M_{2}^{-}$ ($B_{u}$) & $1$ & $-1$ & $1$ & $-1$ \\ \hline
\end{tabular}
\caption{The character table of $C_{2h}$ is shown. 
Each irreducible representation of $C_{2h}$ determines different order parameter (OP) of the CDW coupling $\Gamma$ and $M$.
\label{tab:C2h}}
\end{center}
\end{table}

%\section{Inversion Symmetrization of the system}
%\section{Primary and Secondary Order parameter contribution}
%\section{Mixing contribution from different types of order parameters}
\section{Ginzburg-Landau theory \label{Sec:GLeng}}
The GL free-energy functional is a powerful tool to study phase transitions.  
When a continuous phase transition happens, the symmetry of the system will be broken from a parent space group ($\mathcal{G}_0$) at higher temperature to a lower symmetry subgroup ($\mathcal{G}$; an isotropy subgroup of $\mathcal{G}_0$) at lower temperature.  
Each possible low temperature space group $\mathcal{G}$ has corresponding OPs. 
From the isotopy group analysis accompanying with all possible symmetries of $\mathcal{G}_0$, the functional form of GL free energy is determined and expanded as a function of OPs to some degrees \cite{isotropyG}.
The solution to the GL free energy determines the resulting components of OP.
In our case of interests, the (primary) OP is the triple-$q$ $M_1^-$ CDW mode, denoted by a vector $\vec{\phi} = (\phi_1, \phi_2, \phi_3)$. The $M_1^-$ representation (Table \ref{tab:C2h}) delineated by 
 \begin{equation}
 	M_1^- = A_u \otimes (e^{-i\mathbf{q}_1\cdot \mathbf{T}}, 
 	    e^{-i\mathbf{q}_2\cdot \mathbf{T}}, 
 		e^{-i\mathbf{q}_3\cdot \mathbf{T}}), 
 \end{equation}
where $A_u$ is the IR of $C_{2h}$ for each M pointed by ordering vectors $\{\mathbf{q}_1, \mathbf{q}_2, \mathbf{q}_3\}$ [Fig.\ \ref{fig:2} (c)] and $e^{-i\mathbf{q}_i\cdot \mathbf{T}}$ describes the IR from translation $\mathbf{T}$.
Therefore, the images of $\vec{\phi}$ under symmetry operation are as follows:
\begin{align}
    m_{[100]} &:& (\phi_1, \phi_2, \phi_3) &\rightarrow (-\phi_3, -\phi_2, -\phi_1),    \nonumber  \\    
    3_{[001]}    &:& (\phi_1, \phi_2, \phi_3) &\rightarrow (\phi_2, \phi_3, \phi_1),   \nonumber     \\
    i            &:& (\phi_1, \phi_2, \phi_3) &\rightarrow (-\phi_1, -\phi_2, -\phi_3),   \label{Eq:phi_Tr}   \\    
    T_{\mathbf{a}_1}       &:& (\phi_1, \phi_2, \phi_3) &\rightarrow (-\phi_1, \phi_2, -\phi_3),   \nonumber   \\
    T_{\mathbf{a}_2}      &:& (\phi_1, \phi_2, \phi_3) &\rightarrow (-\phi_1, -\phi_2, \phi_3),   \nonumber
\end{align}
where $T_{\mathbf{a}_1}$ ($T_{\mathbf{a}_2}$) are translation along the primitive lattice vectors $\mathbf{a}_1$ ($\mathbf{a}_2$) in the normal phase.
Following the transformation, the GL free energy of the $M_1^-$ OP near the phase transition temperature has the form of $F_{p} = \frac{1}{2}\alpha\vec{\phi}^2+\frac{1}{4}\beta_1 \vec{\phi}^4 + \frac{1}{4}\beta_2(\phi_1^4+ \phi_2^4 + \phi_3^4)$, where $\vec{\phi}^2 = \phi_1^2 + \phi_2^2 + \phi_3^2$ and $\vec{\phi}^4 = (\vec{\phi}^2)^2$. 
Due to conservation laws, when two $M_1^-$ components of different $\mathbf{q}$s, say $\mathbf{q}_1$ and $\mathbf{q}_2$, couple, they will produce a mode of $\mathbf{q}_3$ with different symmetries, Fig.~\ref{fig:1}(d). The resulting mode turns out to take the $M_1^+$ representation, denoted by
 \begin{equation}
 	M_1^+ = A_g \otimes (e^{-i\mathbf{q}_1\cdot \mathbf{T}}, 
 	    e^{-i\mathbf{q}_2\cdot \mathbf{T}}, 
 		e^{-i\mathbf{q}_3\cdot \mathbf{T}}). 
 \end{equation}
These inevitable $M_1^+$ modes will be collectively called a secondary OP with three components $\vec{\xi}=(\xi_1, \xi_2, \xi_3)$. 
The secondary OP will raise the GL free energy by $F_{s} = \frac{1}{2}\alpha^\prime\vec{\xi}^2+\frac{1}{4}\beta_1^\prime \vec{\xi}^4 + \frac{1}{4}\beta'_2(\xi_1^4+ \xi_2^4 + \xi_3^4) + \gamma' \xi_1 \xi_2 \xi_3$ and by $F_{c} = \lambda (\phi_1 \phi_2 \xi_3 + \phi_2 \phi_3 \xi_1 + \phi_3 \phi_1 \xi_2)$ for interaction with the primary OP.
Since the transformation of the $M_1^+$ OP is the same as Eq.\ (\ref{Eq:phi_Tr}) except the following two,
\begin{align}
    m_{[100]} &:& (\xi_1, \xi_2, \xi_3) &\rightarrow (\xi_3, \xi_2, \xi_1),    \nonumber  \\    
    i           &:& (\xi_1, \xi_2, \xi_3) &\rightarrow (\xi_1, \xi_2, \xi_3),   \label{Eq:xi_Tr}   
\end{align}
it is clear that the $F_{s}$ and $F_{c}$ terms are invariant under the symmetry operations of $G_0$. 
The overall GL energy will be $F_{p}  + F_{s} + F_{c}$. Under the assumption that the primary CDW occurs under the critical temperature and dominates the phase transition ($F_{s}$ is negligible), nonzero $\vec{\phi}$ always induces nonzero $\vec{\xi}$ via the interaction term $F_{c}$ regardless the sign of $\lambda$, indicating the coexistence of $M_1^-$ and $M_1^+$ CDWs.

Essentially, a CDW produces CDW gaps $\Delta$ to the electronic system (multiple gaps for multiple band hybridization). The gaps take symmetries according to the symmetries of the CDW OP. However, the symmetries of the gaps are not necessarily identical to those of the CDW OP. Instead, the former depends on which two bands are involved and the IRs (symmetries) of the associated bands play a role. To understand this, we show the electron-phonon interaction in second quantization
\begin{equation}
H_{\mathrm{el-ph}} \sim \sum_{\mathbf{k,q}} g_{\mathbf{k,q}} c_{\mathbf{k}}^{\dagger} c_{\mathbf{k-q}} \left(b_{\mathbf{q}} + b_{\mathbf{-q}}^{\dagger} \right)
\end{equation}
 for an electron ($c_{\mathbf{k-q}}$) absorbs or emits a phonon ($b_{\mathbf{q}}$, $b_{\mathbf{-q}}^{\dagger}$) into another electron at different momentum ($c_{\mathbf{k}}^{\dagger}$).
A CDW of an ordering wave vectors $\mathbf{q}$ appears when the phonon condensates $\langle b_{\mathbf{q}} \rangle  \neq 0$. (If readers do not take the lattice vibrations as the driving force of the CDW, they can substitute collective charge excitations for the phonons.) After condensation, the ``folded" band electron of momentum $\mathbf{k}$ is defined to be $d_{\mathbf{k}} = c_{\mathbf{k-q}} \langle b_{\mathbf{q} \rangle+ b_{\mathbf{-q}}^{\dagger} }$ and its symmetries are characterized by the symmetries of the ``unfolded" band electron $c_{\mathbf{k-q}}$ and those of the CDW OP $ \langle b_{\mathbf{q}}+ b_{\mathbf{-q}}^{\dagger}  \rangle$. As the result, the IR of the folded electron is the direct product as 
\begin{equation}
\Gamma^{\mathbf{k}}_{\mathrm{folded}} = \Gamma^{\mathbf{k-q}}_{\mathrm{unfolded}} \otimes \Gamma^{\mathbf{q}}_{\mathrm{CDW}}, \label{irrep_band}
\end{equation} 
which was pointed out in our previous work~\cite{huang2018}. Consequently, the CDW gap between CDW bands $a$ and $b$ takes the symmetries of the IR 
\begin{equation}
\Gamma^{\mathbf{k}}_{a} \otimes \Gamma^{\mathbf{k}}_{b}=\Gamma^{\mathbf{k}}_{\mathrm{unfolded-}a} \otimes \Gamma^{\mathbf{k-q}}_{\mathrm{unfolded-} b} \otimes \Gamma^{\mathbf{q}}_{\mathrm{CDW}}.
\label{irrep_gap}
\end{equation}

% ------------------------- %
%              figure       %
% ------------------------- %

%\section{Results for decoupling and mixing types of coupling}
%\section{Results from symmetrization and mixing of OP}
\section{Parametrized Hamiltonian \label{Sec:Mixing_TB}}
To study effects of the CDW, it is necessary to understand the modification of the Hamiltonian from the CDW. The CDW Hamiltonian is written as $H= H_0 + H'$, where $H_0$ is the normal-state Hamiltonian and $H'$ is the perturbation from the CDW. To have translation symmetry and a direct comparison, both $H_0$ and $H'$ are prepared in a basis of $2\times 2$ supercell. $H_0$ falls in space group $P\bar{3}m1$ and $H$ is in space group $P321$. The former contains inversion symmetry, while the latter does not due to the presence of the $M_1^-$ mode that is not invariant under inversion. (The $M_1^+$ mode is inversion-symmetric.) Technically, we can decompose $H'$ into two terms
\begin{equation}
H'  = h_{e} + h_{o}
\end{equation}
by assigning the $M_1^+$ mode to $h_e$ and the $M_1^-$ mode to $h_o$, respectively. The method is depicted as follows.
Suppose a representation of inversion operation $I$ in the normal state $H_0$ is chosen, giving $I H_0 I^{-1} = H_0$ for inversion symmetry. 
The CDW perturbation $H'$, instead, gives $I H' I^{-1} \neq H'$ for breaking inversion symmetry no matter what choice of $I$ is.
Since the $M_1^+$ mode ($M_1^-$) preserves (breaks) inversion symmetry, with an appropriate choice of $I$, we can claim $I h_e I^{-1} = h_e$ ($I h_o I^{-1} = -h_o$). As the result, with $H'$ and $I$, we define 
\begin{align}
\begin{split}
h_{e}   &= \frac{1}{2}\left(H' + I H' I^{-1}\right) ,   \label{Eq:he_ho}  \\
h_{o}   &= \frac{1}{2}\left(H' - I H' I^{-1}\right).    % \nonumber 
\end{split}
\end{align}
The procedure to symmetrize and anti-symmetrize the Hamiltonian can also refer to Refs. \onlinecite{TBsymmerize07, TBsymmerize18}; the same idea for other symmetries is applicable.
We note that the symmetrization depends on the choice of the inversion center (representation of $I$); different choices result in different $h_{e}$ and $h_{o}$. Here we choose the inversion center at a Ti site.

%$H^{\mu \nu}(\mathbf{R}_{ij}) \equiv \langle \mathbf{R}_i, \mu | H | \mathbf{R}_j, \nu \rangle$ between Wannier orbtial-$\mu$ located at cell $\mathbf{R}_i$ and orbital-$\nu$ at cell $\mathbf{R}_j$, where $\mathbf{R}_{ij} = \mathbf{R}_i - \mathbf{R}_i$. 
%Considering a point group operation $g$ with inverse $g^{-1}$, the transformation of the basis is defined as follows
%\begin{align}
%g^{-1}: &~ 	 \mu,~\nu  \longmapsto \mu',~\nu'    \nonumber	\\
%	&~ \mathbf{R}_i,~ \mathbf{R}_j 	 \longmapsto \mathbf{R}_i'=g^{-1} \mathbf{R}_i,~ \mathbf{R}_j'=g^{-1} \mathbf{R}_j. 
%	  \label{Eq:g_ti}
%\end{align}
%As the Hamiltonian $H$ is invariant under $g$, $H^{ij}(\mathbf{R}_{ij})$ can then be related to the transformed basis as 
%\begin{equation}
%H^{\mu \nu}(\mathbf{R}_{ij}) = \sum_{\mu' \nu'} D^{\mu \mu'}(g)H^{\mu' \nu'}(\mathbf{R}_{ij}')  D^{\nu' \nu}(g^{-1}),
%\end{equation} 
%where $D^{\mu \mu'}(g)$ is the representation of $g$ between orbital $\mu$ and $\mu'$, and similarly for $D^{\nu' \nu}(g^{-1})$. 

Once $H'$ is decomposed, we study the effects of $h_{e}$ and $h_{o}$ respectively.
We parametrize the Hamiltonian as 
\begin{equation}
H_{\left(\alpha, \beta \right)} = H_0 + \alpha  h_{e} + \beta  h_{o}
\label{Eq:H_alpha_beta}
\end{equation}
with $(\alpha, \beta)$ as the tuning parameters to control strengths of CDW OPs of different parity.
In the following subsections, we will demonstrate $\mathbb{Z}_2$-topologies of parametrized CDW states for purified ($\alpha \beta=0$) and mixed ($\alpha \beta \neq 0$) Hamiltonians, respectively, by calculating the Wilson loop (Wannier charge center) \cite{Wilson11_Vanderbilt, Wilson11_XDai}. 
Since the $M_1^-$ mode is primary and the $M_1^+$ mode is secondary, in our construction the strength of the even-parity $h_e$ is much smaller than the odd-parity $h_o$. To have comparable contribution from the two perturbations, $\mathcal{O}(\alpha)>\mathcal{O}(\beta)$ will be taken.

\subsection{Purified $H$}
A purified $H$ is defined for only one of $h_{e} $ and $h_{o}$ existing. 
Starting with the system $H_{\left(0, 1\right)} = H_0 + h_{o}$ for an artificial CDW state with the $M_{1}^{-}$ mode but without $M_{1}^{+}$, we find that a CDW gap opens at the Fermi energy but its $\mathbb{Z}_2$ topology is trivial. 
To test whether the topology remains trivial, we tune the value of $\beta$ as lower as $0.05$ and we do not find a topological phase transition. 
The band structure of the system $H_{\left(0, 0.2 \right)}$ having a small gap is shown in Fig.\ \ref{fig:3} (d).
We interpret that the $\alpha=\beta=0$ case is an unstable nodal-ring semimetal phase, rings without topology, and the nodal rings are gapped once the CDW is on.

In contrast, the purely parity-even system $H_{\left(1, 0\right)} = H_0 + h_{e}$ has an almost vanishing CDW gap since $h_{e}$ is small as an outcome of the secondary OP. 
The system has no clear gap so that its topological aspect cannot be determined easily unless the opened gap is large enough. 
To have an appreciable gap for clear identification of topology, the situation of enlarged $\alpha$ is discussed. 
With $\alpha$ ranging from 3 to 7.3, the gap is visible; the gap at $\alpha=4$ is still very small [Fig.\ \ref{fig:4} (a)] and a larger one at $\alpha=7$ is seen [Fig.\ \ref{fig:4} (b)].
The Wilson loop calculations show them all to be topologically non-trivial
as seen in Fig.~\ref{fig:4} (e-f) where an odd number of crossing points on a horizontal line along the $k_y$ direction.

We summarize the findings here that in the case of either a purely parity-even or a purely parity-odd CDW perturbation, the strength of the OP only modifies the band structure but does not change the topology. The purely parity-even and parity-odd CDW states are topologically nontrivial ($\nu=1$) and trivial ($\nu=0$), respectively.

%\subsection{Mixing can have different topologies}
\subsection{Mixed $H$}

\begin{figure*}
\begin{center}
\includegraphics[width=\textwidth]{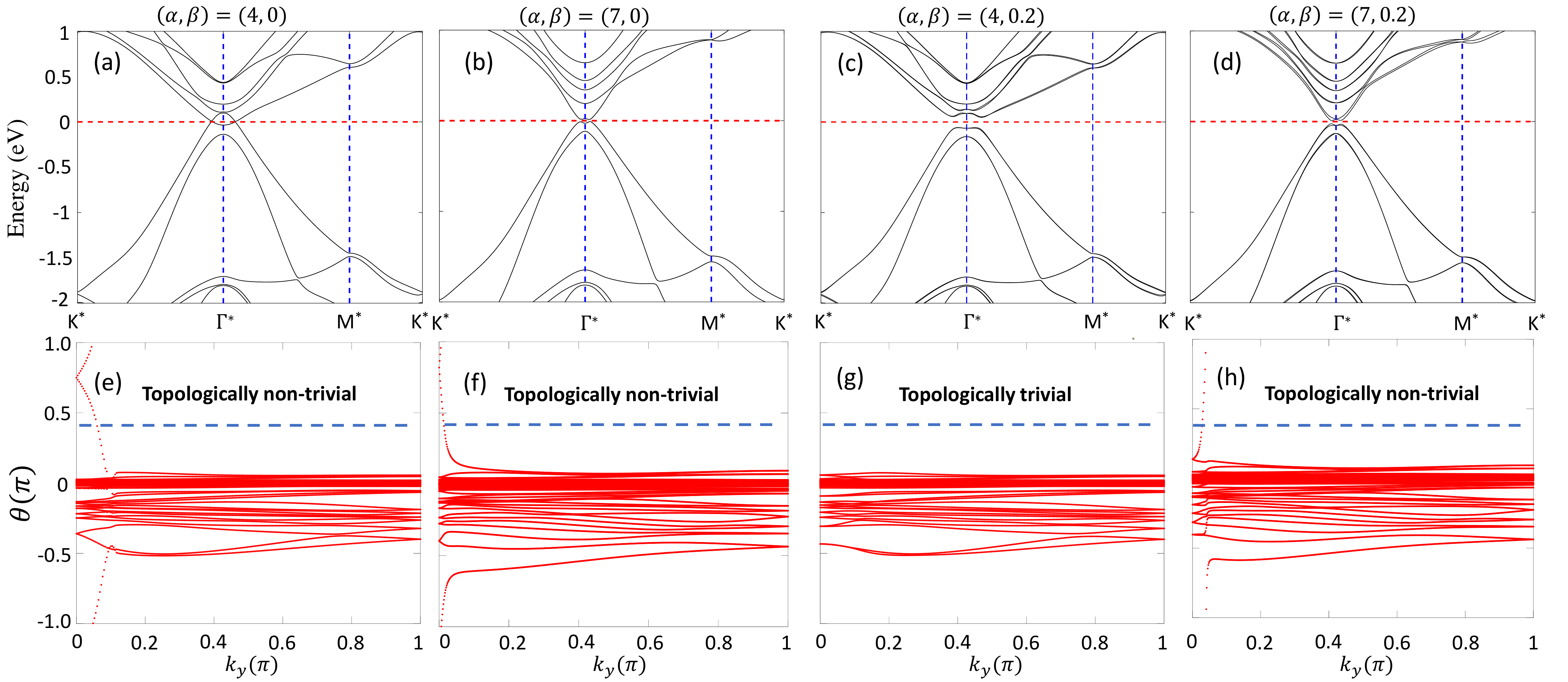}
\caption{
Band structures of the CDW states with different tuning parameters $(\alpha, \beta)$ in (a-d). The corresponding Wilson loop calculations are shown in (e-h).
\label{fig:4}}
\end{center}
\end{figure*}

Due to opposite topological tendencies resulted from $h_e$ and $h_o$ respectively, change of the system topology is possible via mixing them by tuning $(\alpha, \beta)$.
The conclusive topological phase diagram is shown in Fig.~\ref{fig:5}, where a blurred phase boundary is due to the uncertain vanishing band gap beyond our numerical capacity.
The original CDW Hamiltonian from the DFT-HSE calculation, $H_{\left(1, 1\right)}$, shows trivial topology.
Referring to the previous findings, it is expected that $H_{\left(\alpha, \beta\right)}$ tends to become topologically non-trivial when $\alpha \gg \beta$. 
With fixed $\beta$, the system $H_{\left(\alpha, \beta\right)}$ become topologically non-trivial when $\alpha$ is increased.
For example, comparing two band insulators, $H_{\left(4, 0.2\right)}$ [Fig.\ \ref{fig:4} (c)] and $H_{\left(7, 0.2\right)}$ [Fig.\ \ref{fig:4} (d)], the former is topologically trivial [Fig.\ \ref{fig:4} (g)], whereas the latter is topologically non-trivial [Fig.\ \ref{fig:4} (h)].
Therefore, we demonstrate a topological phase transition realized by tuning a parameter that controls the inversion breaking effect~\cite{Murakami2008}.

\section{Discussion on Topology \label{Sec:discussion}}

In the primitive unit cell the normal state $H_0$ has the band structure shown in Fig. \ref{fig:3} (a), while the bands are folded into the RBZ when $H_0$ is described by a $2 \times 2$ supercell [Fig.\ \ref{fig:3} (c)].
The bands from the $M$ points in the primitive BZ become three-fold degeneracies at the $\Gamma^*$ point in the RBZ. 
Therefore, the supercell has the RBZ band structure from one single top valence band $V_\Gamma$ at $\Gamma$ band inverting with three bottom conduction bands $C_M$s at $M$ in the primitive BZ.
Because of $\Gamma$ and $M$s being inversion-invariant points, $V_\Gamma$ is found to be characterized by parity $-1$ and $C_M$ by parity $+1$ [Fig.\ \ref{fig:3} (a)].
Some questions will be raised like, ``How to characterize these folded bands?" and ``Is the band inversion topologically nontrivial or not?" 
Before answering them, we reduce the system of interest to a two-band model by taking the lowest-energy band of the three folded bands. Only the lowest one is considered because its band inversion with $V_\Gamma$ determined the topology of the CDW state. According to group theory, the three folded bands will hybridize into one singlet and one doublet states at $\Gamma^*$ and the doublet is further split into two by SOC. (At this moment assume that inversion symmetry is still present and thus each band is doubly degenerate from two spins.) We are not concerned about which one is of the lowest energy; parity of three bands at $\Gamma^*$ is identical. In the following, we will  denote the lowest folded one by $C'_M$.

To construct an effective model of the two orbitals (bands) $\{ V_\Gamma, C'_M\}$ and two spins around $\Gamma^*$, we will use Pauli matrices $\tau$s and $\sigma$s acting in orbital and spin spaces, respectively. Identity matrices $\tau_0$ and $\sigma_0$ are also taken.
The Hamiltonian will take the form
\begin{equation} \label{BHZ}
\mathcal{H}(\mathbf{k}) = \hat{\epsilon}(\mathbf{k}) \tau_z + \hat{\Delta}(\mathbf{k})\tau_+ + \hat{\Delta}^{\dagger}(\mathbf{k})\tau_-,
\end{equation}
where a diagonal term of $\tau_0$ is omitted without affecting the topology. $\hat{\epsilon}(\mathbf{k}) \tau_z$ discribes the dispersions of the hole and folded electron pockets with opposite curvatures by taking
\begin{equation}
\hat{\epsilon}(\mathbf{k}) = (m-A k^2) \sigma_0 + \cdots
\end{equation}
with $m$ and $A$ being positive accounting for an energy overlap and with $\cdots$ for contributions from SOC. The off-diagonal term $\tau_{\pm}=\left(\tau_{x}\pm i\tau_y \right)/2$ captures a CDW gap function $\hat{\Delta}(\mathbf{k})$, which is a $2\times 2$ matrix for spin degrees of freedom.

When the Hamiltonian at $\Gamma^*$ is invariant under the operations in the symmetry group, the IR (or symmetry) of the gap function will be the direct product of the IRs of $V_\Gamma$ and $C'_M$ as in Eq.~(\ref{irrep_gap}). Consequently, under inversion, the Hamiltonian is transformed as 
$\mathcal{I} \mathcal{H}(\mathbf{k}) \mathcal{I}^{-1} = \mathcal{H}(-\mathbf{k})$, indicating that the CDW gap function between $V_\Gamma$ and $C'_M$ follows $ \hat{\Delta}(-\mathbf{k}) = \eta \hat{\Delta}(\mathbf{k})$,
where $\eta=\pm1$ according to the IR of the gap function. 
The parity of the folded $C'_M$ is $(+1)\times(+1)=+1$ via the $M_1^+$ mode, one $+1$ from the character of $C_M$ and the other $+1$ from that of the CDW OP. Recalling the parity $-1$ of $V_\Gamma$, the inversion operator is therefore written as $\mathcal{I} = \mathcal{I}_{M_1^+} = -\tau_z \sigma_0$ in this basis ($\sigma_0$ for no effect on the spin under inversion). The subscript is to mark that the $C'_M$ and $C_M$ are associated by the $M_1^+$ mode. As the result, the gap function is odd in $\mathbf{k}$, $\hat{\Delta}_{M_1^{+}}(-\mathbf{k})  = - \hat{\Delta}_{M_1^{+}}(\mathbf{k})$. 
When time-reversal symmetry is also considered, the Hamiltonian satisfies $\Theta  \mathcal{H}(\mathbf{k}) \Theta^{-1} = \mathcal{H}(-\mathbf{k})$, where $\Theta=i\sigma_y K$, making $\Theta \vec{\sigma} \Theta^{-1} = -\vec{\sigma}$.
A possible choice of $\hat{\Delta}_{M_1^{+}}(-\mathbf{k})$, to leading order of $\mathbf{k}$, will be: 
\begin{equation}
\hat{\Delta}_{M_1^{+}}(\mathbf{k}) \approx -i\Delta_{+} k_x \left( k_x^2-3k_y^2\right)\sigma_0 +\Delta_{\text{S}} \left( k_y \sigma_x - k_x \sigma_y \right),  \label{Eq:gap_Mp}
\end{equation}
($\Delta_{+},~\Delta_{\text{S}}\in \mathbb{R}$) to retain the rotation symmetry and mirror symmetry $m_{[100]}$. $\Delta_{+} k_x \left( k_x^2-3k_y^2\right)$ is the CDW gap without SOC, which reflects the oddness relative to three mirror planes. The $\Delta_{\text{S}}$ term is the emergent Rashba-like SOC, which preserves the rotation symmetry. 
Evidently, the resulting Hamiltonian $\mathcal{H}(\mathbf{k})$ is to model a $\mathbb{Z}_2$ topological insulator.
By transforming $\tau_x \rightarrow \tau_y \sigma_x,~\tau_y \rightarrow -\tau_x \sigma_x,~\tau_z \rightarrow \tau_z$ and $\sigma_x \rightarrow \sigma_x,~\sigma_y \rightarrow \tau_z \sigma_y,~\sigma_z \rightarrow \tau_z \sigma_z$ in Eq. (\ref{BHZ}) and redefining the spin coordinates, the Hamiltonian becomes the 2D Bernevig-Hughes-Zhang (BHZ) model~\cite{BHZ06} as $\mathcal{H}(\mathbf{k}) =  (m-A k^2) \sigma_0 \tau_z + \Delta'_{\text{S}} k_x \sigma_z \tau_x - \Delta_{\text{S}} k_y \sigma_0 \tau_y $, where $\Delta'_{\text{S}} = \sqrt{\Delta_{\text{S}}^2+\Delta_{+}^2\left( k_x^2-3k_y^2\right)^2}$.

\begin{figure}
\begin{center}
\includegraphics[width=0.5\textwidth]{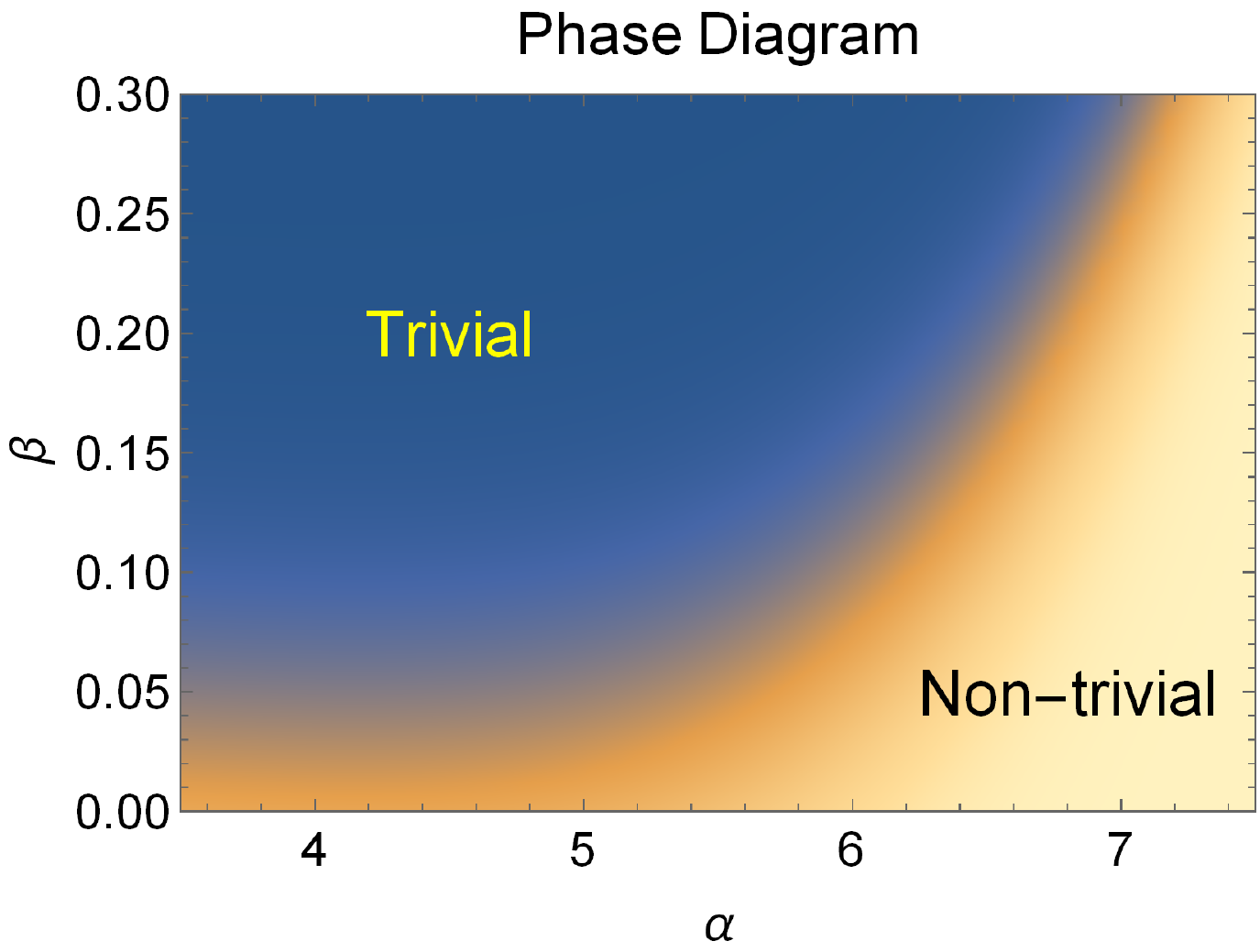}
\caption{
The topological phase diagram with respective to the tuning parameter $(\alpha, \beta)$ from the Wilson loop calculation is shown.
The blurred region is the estimated boundary of the two phases, since near the vanishing gap region the Wilson loop calculation is not an appropriate method to determine the topology.
\label{fig:5}}
\end{center}
\end{figure}

\begin{figure}
\begin{center}
\includegraphics[width=0.5\textwidth]{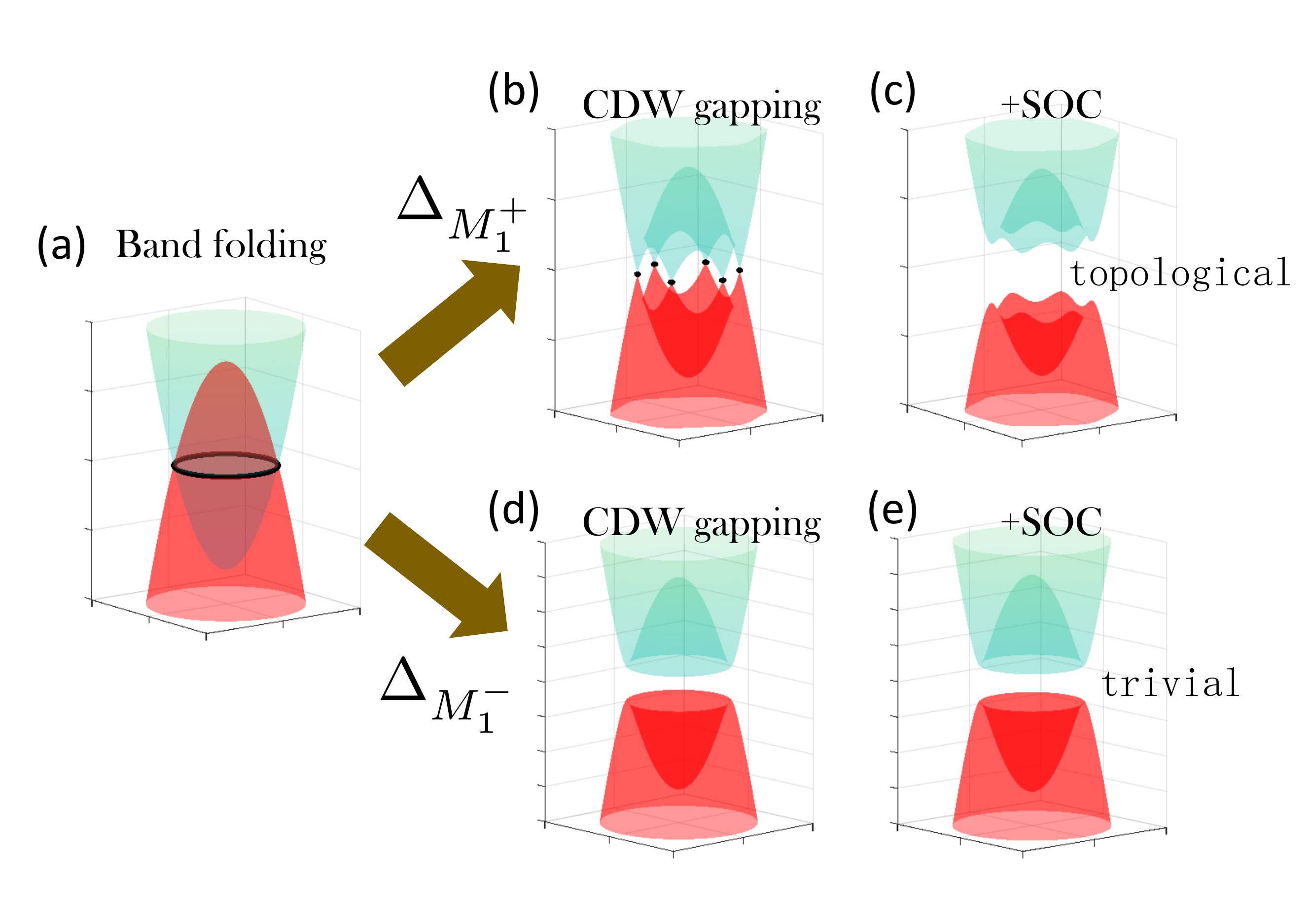}
\caption{
(a) The nodal ring formed from the band-folding of the supercell Hamiltonian.
(b), (c) The band structure perturbed by the CDW $M_1^+$ gap function without and with SOC. (d) and (e) account for the band structures with the $M_1^-$ gap without and with SOC, respectively. 
\label{fig:6}}
\end{center} 
\end{figure}

However, inversion symmetry should not exist when the $M_1^{-}$ mode is considered. 
Like the idea before, we add a term for the $M_1^{-}$ mode, $\hat{\Delta}_{M_1^{-}}(\mathbf{k})$, to the gap function,  such that the inversion symmetry is lost. To this end, $\hat{\Delta}_{M_1^{-}}(\mathbf{k})$ takes different parity from that of $\hat{\Delta}_{M_1^{+}}(\mathbf{k})$, and then $\hat{\Delta}_{M_1^{-}}(-\mathbf{k})  = \hat{\Delta}_{M_1^{-}}(\mathbf{k})$. A suggested gap function is
\begin{equation}
\hat{\Delta}_{M_1^{-}}(\mathbf{k}) \approx  \Delta_{-}  \sigma_0. \label{Eq:gap_Mm}
\end{equation} 
Meanwhile, a Rashba SOC can contribute to $\hat{\epsilon}(\mathbf{k})$, which will not change the topology in our study.
We understand that the addition of $\hat{\Delta}_{M_1^{-}}(\mathbf{k})$ is a disadvantage to the topology because it hinders the ``winding" of the spin texture.
The spin texture is disclosed in the eigenvalue of of $\hat{\Delta}(\mathbf{k})  = \hat{\Delta}_{M_1^{+}}(\mathbf{k}) +\hat{\Delta}_{M_1^{-}}(\mathbf{k})$, that is $\Delta_{-}\pm\Delta_{+}|\mathbf{k}|$ by setting $\Delta_{\text{S}}=0$ for simplicity. 
When the $M_1^-$ dominates, i.e. large $\Delta_-$, the eigenvalue $\Delta_{-}\pm\Delta_{+}|\mathbf{k}|$ does not change sign when $\mathbf{k}$ runs over the BZ.
Therefore, the system is estimated to be topologically trivial when $|\Delta_{-}| \gtrsim |\Delta_{+}| \pi$ (over a superlattice constant).
This explanation answers the observation in Fig.~\ref{fig:5} that the $M_1^{+}$ mode drives to the topological state, while the $M_1^{-}$ mode makes trivial.
Figure \ref{fig:6} shows a schematic to demonstrate the band structures of two CDW gaps. Two CDWs, $M_1^{+}$ and $M_1^{-}$, show distinct gap functions. Starting with the band-folding, the conduction and valence bands cross as a nodal ring, which is fragile against any perturbation. When the interband coupling is considered, in the absence of SOC, the nodal line turns into six isolated Dirac nodes for the $M_1^{+}$ CDW case [due to the first term in Eq. (\ref{Eq:gap_Mp})], but it is completely gapped for the $M_1^{-}$ CDW case [due to Eq. (\ref{Eq:gap_Mm})]. Only the former band structure can become a topological insulator when SOC is on.

We provide another understanding by attributing the folded $C'_M$ to the $M_1^-$ mode to account for its primary role. In consequence, the inversion operation will take
$\mathcal{I} = \mathcal{I}_{M_1^-} = -\tau_0 \sigma_0$ that works \textit{locally} for $\{ V_\Gamma, C'_M\}$. The $M_1^+$ mode will be taken as a perturbation to break the equality $\mathcal{I} \mathcal{H}(\mathbf{k}) \mathcal{I}^{-1} = \mathcal{H}(-\mathbf{k})$. This choice of $\mathcal{I}$ will lead to the same form of the gap functions as in Eqs.~(\ref{Eq:gap_Mp}) and (\ref{Eq:gap_Mm}), which, however, originate differently 
that $\hat{\Delta}_{M_1^{+}}$ preserves inversion symmetry while $\hat{\Delta}_{M_1^{-}}$ breaks. Clearly, a CDW state starting with the gap function $\hat{\Delta}_{M_1^{-}}(\mathbf{k})$ in Eq.~(\ref{Eq:gap_Mm}), a typical CDW gap, is a trivial insulator.

% ------------------------- %
%              figure       %
% ------------------------- %

\section{Conclusion}
A typical topological phase transition is realized when a direct gap between the valence band and the conduction bands at some (or multiple) momentum $k$ closes, resulting in a band touching and hence a band inversion. A different scenario that a band inversion occurs between states at distinct $k$ points becomes possible when translation symmetry is broken by a CDW. In such case, the orbital characters in two inverted states are crucial to affect the topology of the system. The characters are elucidated in Eq. (\ref{irrep_band}) and Ref. \cite{huang2018}. Monolayer 1\textit{T}-TiSe$_2$ perfectly exemplifies the story for its CDW produces an indirect band inversion between the hole pocket at $\Gamma$ and electron pockets at $M$s (for the bulk compound the electron pockets reside at $L$s). Monolayer 1\textit{T}-TiSe$_2$ is a little tricky than the bulk compound because the primary CDW is a triple-$q$ $M_1^-$ mode and it will generate the triple-$q$ $M_1^+$ mode, such that inversion symmetry disappears and two modes are mixed all the time.
The CDW of different symmetries can affect the system electronic structure and topology quite oppositely, so they have competitive roles. This competition for distinct topologies is worth studying.

To study the effect of CDW of different symmetries on topology, we symmetrize the perturbation with respect to inversion for the tight-binding (TB) model obtained from the wannierization method applying for both the primitive phase and the CDW phase.
The unperturbed Hamiltonian is the supercell with band inversion between electron and hole pockets having positive and negative parity respectively.
The topology of the system with purely parity-even perturbations is found to be topologically non-trivial, while the system with purely parity-odd perturbation is topologically trivial.
Via tuning strengths of parity-even and parity-odd perturbations, a topological phase transition can be achieved.
These results are explained sufficiently by an effective four-band model with a parity-mixed CDW gap function.

\begin{acknowledgments}
S.M.H is supported by the Ministry of Science and Technology (MoST) in Taiwan under grant No. MOST 108-2112-M-110-013-MY3.
H.L. acknowledges the support by the Ministry of Science and Technology (MOST) in Taiwan under grant number MOST 109-2112-M-001-014-MY3.
\end{acknowledgments}
\section{Appendix}

In a purely parity-even perturbed system, we can use parity products to discuss its topology.
We have take the fact that the parities at the $M^*$ points in the RBZ have no effects on the topology so as to be neglected in the above discussion. 
This is because each folded primitive band is split into two at $M^*$ (also at the RBZ boundaries) with one of even parity and the other of odd parity. 
Besides, the original primitive valence bands are of even number so that their parity product at the $M^*$ point will contribute no change to the $\mathbb{Z}_2$ invariant~\cite{huang2018}. The proof is as follows. 
For the normal state of $H_0$, the band structure in the primitive BZ is obtained from $H_0\psi_0(k) = \epsilon_0(k) \psi_0(k)$, where $\psi_0(k)$ is the eigenstate and the eigenvalue $\epsilon_0(k) = \epsilon_0(-k)$ since $H_0$ is inversion symmetric.
Considering the supercell, the $M^*$-point state in the RBZ is folded from $k_1=M^*$ and $k_2=-M^*$ in the primitive BZ, which are degenerate in energy [$\epsilon_0(k_1) = \epsilon_0(k_2) \equiv \varepsilon_{p, M^*}$].
The added perturbation $h_{e}$ obeys the inversion symmetry so that it introduces coupling between them satisfying
\[  \delta \equiv  \langle \psi_0(k_2) | h_{e} |  \psi_0(k_1) \rangle = \langle \psi_0(-k_2) | h_{e} | \psi_0(-k_1) \rangle,
\] 
where $I \psi_0(k) = \psi_0(-k)$ for some gauge. 
Since $k_1=M^*$ and $k_2=-M^*$ are inversion symmetric to each other, this results in the coupling $\delta=\delta^*$ being a real number.
Therefore, the perturbed system $H_{\left(\alpha, 0\right)}=H_0+\alpha h_e$ at the $M^*$ point can be solved in the basis of  $(\psi_0(k_1), \psi_0(k_2))^T$ as
\begin{equation}
    \left(  \begin{array}{cc}
              \varepsilon_{p, M^*}   &   \alpha  \delta            \\
              \alpha \delta           &    \varepsilon_{p, M^*}     \\
               \end{array}
               \right).
\end{equation}
Therefore, each folded band at $M^*$ is split into two eigenstates $\psi^{e,o} = (\psi_0(k_1) \pm \psi_0(k_2))/\sqrt{2}$ for even and odd parity respectively.

\end{document}